\documentclass{ws-ijmpa}

\newcommand{\be}{\begin{equation}}
\newcommand{\ee}{\end{equation}}
\newcommand{\ba}{\begin{eqnarray}}
\newcommand{\ea}{\end{eqnarray}}

\def\gl#1{(\ref{#1})}

\begin{document}

\markboth{A. A. Andrianov, V. A. Andrianov}{Quasilocal Quark Models as
Effective Theory}
\catchline{}{}{}{}{}

\title{Quasilocal Quark Models as Effective Theory of
Non-perturbative QCD\footnote{Talk at the International Symposium
"MENU 2004", 29.August-4.September, Beijing,China.}}

\author{A. A. Andrianov}
\address{Dipartimento di Fisica, Universita'
di Bologna and Istituto Nazionale di Fisica Nucleare,
Sezione di Bologna, 40126 Bologna, Italia, andrianov@bo.infn.it}

\author{V. A. Andrianov}
\address{V.A.Fock Department of Theoretical
Physics, St.Petersburg State University,\\
198504 St.Petersburg, Russia, Vladimir.Andrianov@pobox.spbu.ru}

\maketitle


\begin{abstract}
We consider the Quasilocal Quark Model of NJL type (QNJLM) as an effective theory
of non-perturbative QCD including scalar (S), pseudoscalar (P), vector (V) and
axial-vector (A) four-fermion interaction with derivatives. In the presence of
a strong attraction in the scalar channel the chiral symmetry
is spontaneously broken
and as a consequence the composite meson states are  generated in all
channels. With the help of Operator Product Expansion the appropriate set
of Chiral Symmetry Restoration (CSR) Sum Rules
in these channels are imposed as matching conditions to QCD at intermediate
energies. The mass spectrum and some decay constants for ground
and excited  meson states are calculated.
\end{abstract}


\section{Introduction}

The QCD-inspired quark models with four-fermion
interaction are often applied for the effective description of
low-energy QCD in the hadronization regime. The local four-fermion
interaction is involved to induce the dynamical chiral symmetry breaking
(DCSB) due to strong attraction in the scalar channel. As a consequence,
the dynamical quark mass $m_{dyn}$ is created, as well as an isospin
multiplet of pions, massless in the chiral limit, and a massive scalar
meson with mass $m_{\sigma}=2m_{dyn}$ arises. However it is known from
the experiment \cite{eid} that there are series of meson states
with equal quantum numbers and heavier masses, in particular,
$0^{-+}(\pi,\pi',\pi'',...)$; $0^{++}(\sigma
(f_{0}),\sigma',\sigma'',...)$; $1^{--}(\rho,\rho',\rho'',...)$.
Due to confinement, one expects an infinite number of such excited
states with increasing masses. Therefore in order to describe the physics
of those resonances at intermediate energies one can extend the quark model
with local interaction of the Nambu-Jona-Lasinio (NJL) type \cite{njl}
taking into account  higher-dimensional quark operators
with derivatives, i.e. quasilocal quark interactions\cite{[3],[4],[5],[6],[7],zap,[9]}. For
sufficiently strong couplings the new
operators promote formation of additional new meson states.
Such a quasilocal approach (see also \cite{[11],[12],[10]}) represents a
systematic extension of the NJL-model towards the complete
effective action of QCD where many-fermion vertices with
derivatives possess the manifest chiral symmetry of
interaction, motivated by the soft momentum expansion of the
high-energy perturbative QCD effective action.

Another idea is to impose CSR Sum Rules at high energies \cite{zap}.
In particular, at intermediate energies the correlators of QNJLM can
be matched to the Operator Product Expansion (OPE) of QCD
correlators \cite{[20]}. This matching realizes the correspondence
to QCD and improves the predicability of QNJLM. It is based on the
large-$N_{c}$ approach which is equivalent to planar QCD.
In this approximation the correlators of color-singlet quark currents are
saturated by infinite number of narrow meson resonances.

On the other hand
the high-energy asymptotic is
provided \cite{[20]} by the perturbation theory of QCD and the OPE due to
asymptotic freedom of QCD. Therefrom the
correlators under discussion increase at large $p^{2}$:
$\Pi^{C}(p^{2})\mid_{p^{2}\rightarrow\infty}\sim
p^{2} \ln\frac{p^{2}}{\mu^{2}}$.
When comparing the two approaches one concludes that the
infinite series of resonances with the same quantum numbers
should exist in order to reproduce the perturbative asymptotic.

Meantime the differences of correlators of opposite-parity currents
rapidly decrease at large momenta \cite{zap,[12]}:
$(\Pi^{P}(p^{2})-\Pi^{S}(p^{2}))_{p^{2}\rightarrow\infty}\equiv
\frac{\Delta_{SP}}{p^{4}}+O(\frac{1}{p^{6}}),$\\ $\Delta_{SP}\simeq
24\pi\alpha_{s}\langle\bar{q}q\rangle^{2} \label{SP}$
and\cite{[9],[11]}:
$(\Pi^{V}(p^{2})-\Pi^{A}(p^{2}))_{p^{2}\rightarrow\infty}\equiv
\frac{\Delta_{VA}}{p^{6}}+O(\frac{1}{p^{8}}),$\\ $ \Delta_{VA}\simeq
-16\pi\alpha_{s}\langle\bar{q}q\rangle^{2} \label{VA}$,
where
$\langle\bar{q}q\rangle$ is  a quark condensate, we have defined for V,A fields
$\Pi_{\mu\nu}^{V,A}(p^{2})\equiv(-\delta_{\mu\nu}p^{2}+p_{\mu}p_{\nu})
\Pi^{V,A}(p^{2})$,
and the vacuum dominance hypothesis\cite{[20]} in the
large-$N_{c}$ limit is adopted.

Therefore the chiral symmetry is restored at high energies and
the two above differences represent  genuine order parameters of
CSB in QCD. As they decrease rapidly at large momenta
one can perform the matching of QCD asymptotic
by means of few lowest lying resonances that gives a number of
constraints from the CSR.
They may be used
both for obtaining some additional bounds on the model parameters
and for calculating of some decay constants (see in \cite{[10],[14]}
and references therein).
In the present talk the QNJLM is presented with two channels where two
pairs of SPVA-mesons are generated. Respectively it is expected to reproduce the lower
part of QCD meson spectrum.

\section{Quasilocal Quark Model of NJL-type}

The minimal $n$-channel lagrangian of the QNJLM has \cite{[4],[5],[6]}
the following form,
$$
L=\bar{q}i\hat{\partial}q+
\frac{1}{4N_{c}\Lambda^{2}}\sum_{k,l=1}^{2}
\left\{a_{kl}[\bar{q}f_{k}q\cdot\bar{q}f_{l}q+
\bar{q}f_{k}i\gamma_{5}q\cdot\bar{q}f_{l}i\gamma_{5}q]\right.
$$
\be
\left.+b_{kl}[\bar{q}f_{k}i\gamma_{\mu}q\cdot\bar{q}f_{l}i\gamma_{\mu}q
+\bar{q}f_{k}i\gamma_{\mu}\gamma_{5}q\cdot\bar{q}f_{l}
i\gamma_{\mu}\gamma_{5}q]\right\} ,
\ee
where
$a_{kl}, b_{kl}$ represent symmetric matrices of real coupling
constants and $f_{k}$ are formfactors.
We will restrict ourselves by the case $n=2$ and
describe the ground meson states and their first excitations only.

The observables should not depend on
the cutoff $\Lambda$. The  scale invariance is achieved with the help
an appropriate prescription of cutoff dependence for effective
coupling constants $a_{kl},b_{kl}$. Namely, we require the
cancelation of quadratic divergences and
parameterize the matrices of coupling constants in the vicinity
of polycritical point as follows:
$8\pi^{2}a_{kl}^{-1}=\delta_{kl}-\frac{\Delta_{kl}}{\Lambda^{2}};
16\pi^{2}b_{kl}^{-1}=\delta_{kl}-\frac{4}{3}
\frac{\bar\Delta_{kl}}{\Lambda^{2}};
\Delta_{kl}, \bar\Delta_{kl}\ll\Lambda^{2}.$
The last inequalities provide the
masses to be essentially less than the cutoff.

The parameters $\Delta_{kl}$ just describe the deviation from a
critical point and determine the physical masses of scalar mesons.
The CSB is generated by the dynamic quark mass function corresponding
to nontrivial v.e.v.'s of scalar fields $\sigma_1,\sigma_2$:\\
$M(\tau) = \sigma_1 f_1(\tau) + \sigma_2 f_1(\tau);\quad M_0 \equiv
M(0) = 2 \sigma_1;\quad
\tau\equiv-\frac{\partial^{2}}{\Lambda^{2}}$.
The physical mass spectrum can be found from solutions of the
corresponding secular equation,
$det(\hat{A}p^{2}+\hat{B})=0;\quad m_{phys}^{2}=-p_{0}^{2}$,
where $\hat{A}$ and $\hat{B}$ represent the kinetic term and the momentum
independent part correspondingly.
Let us display the mass-spectrum for ground meson states and their first
excitations. We introduce the notations,
$\sigma^2\equiv\sigma_{1}^{2}+
\frac{2\sqrt{3}}{3}\sigma_{1}\sigma_{2}+3\sigma_{2}^{2}>0$,
$d\equiv 3\bar\Delta_{11}+2\sqrt{3}\bar\Delta_{12}+\bar\Delta_{22}$,
and take into account the consistency inequalities
$\Delta_{22}<0, \bar\Delta_{22}<0$.
The spectra for scalar and pseudoscalar mesons are:
$m_{\sigma}=4\sigma_{1}=2M_{0}; \, m_{\pi}=0; \,\label{nambu}$
$m_{\pi'}^{2}\simeq-\frac{4}{3}\Delta_{22}+\sigma^2;\quad
m_{\sigma'}^{2}-m_{\pi'}^{2}\simeq 2\sigma^2>0. \label{sipi}$
The spectra for vector and axial-vector mesons are:
$m_{\rho}^{2}\simeq
-\frac{\mbox{\rm det}\bar\Delta_{kl}}{2\bar\Delta_{22}\ln{\frac{\Lambda^{2}}{M_{0}^{2}}}};
\quad m_{a_1}^{2}\simeq m_{\rho}^{2}+6M_{0}^{2};\label{arho}$
$\quad m_{\rho'}^{2}\simeq-\frac{4}{3}\bar\Delta_{22}-\frac{d}{6\ln{\frac{\Lambda^{2}}{M_{0}^{2}}}}-m_{\rho}^{2};
\quad m_{a'_1}^{2}-m_{\rho'}^{2}\simeq\frac{3}{2}(m_{\sigma'}^{2}-m_{\pi'}^{2})
\simeq 3\sigma^2>0.
\label{exc}$
The prime labels everywhere an excited meson state.

We identify $\sigma$ with $f_{0}(400-1200)$, $\sigma'$ with
$f_{0}(1370)$, $\pi'$ with $\pi(1300)$, $\rho$ with $\rho(770)$,
$\rho'$ with $\rho(1450)$ and $a_1$ with $a_{1}(1230)$.
The experimental data \cite{eid} give us:
$m_{\sigma}=400\div1200 \mbox{Mev};\quad
m_{\sigma'}=1200\div1500 \mbox{MeV};\quad
m_{\pi'}=1300\pm 100 \mbox{ MeV};$
$m_{\rho}=770\pm 0.8 \mbox{ MeV};\quad
m_{\rho'}=1465\pm 25 \mbox{ MeV};\quad
m_{a_{1}}=1230\pm 40 \mbox{ MeV}$.
The prediction for the mass of $\sigma$-meson is then
$m_{\sigma} \simeq 800 \mbox{ MeV}, \label{sig}$
which is close to the averaged experimental value.
Furthermore we have the following prediction for the mass of
$a_{1}'$-particle,
$m_{a_{1}'}\cong 1465 \div 1850 \mbox{ MeV}$.
The large range for a possible mass of $a_{1}'$-meson is accounted for by
a big experimental uncertainty for the mass of $\sigma'$ and
$\pi'$ mesons. If we accept the averaged values for them and use the CSR rules, then
$m_{a_{1}'}-m_{\rho'}\approx 30$ MeV. One can confront this value with a phenomenological
estimate, $m_{a_{1}'}= 1640\pm 40 \mbox{ MeV}$ \cite{eid}.

\section{Chiral Symmetry Restoration Sum Rules }

Let us exploit the constraints based on chiral symmetry
restoration in QCD at high energies. Expanding the meson correlators
in
powers of $p^{2}$ one arrives to the CSR Sum Rules. In the
scalar-pseudoscalar case
they read:
$\sum_{n}Z_{n}^{S}-\sum_{n}Z_{n}^{P}=0$;
$\sum_{n}Z_{n}^{S}m_{S,n}^{2}-\sum_{n}Z_{n}^{P}m_{P,n}^{2}=\Delta_{SP}$,
and in the vector-axial-vector \gl{VA} one obtains:
$\sum_{n}Z_{n}^{V}-\sum_{n}Z_{n}^{A}=4f_{\pi}^{2};
\sum_{n}Z_{n}^{V}m_{V,n}^{2}-\sum_{n}Z_{n}^{A}m_{A,n}^{2}=0,$\\
$\sum_{n}Z_{n}^{V}m_{V,n}^{4}-\sum_{n}Z_{n}^{A}m_{A,n}^{4}=\Delta_{VA}$.
The first two relations  are
the famous Weinberg Sum Rules, with $f_{\pi}$  being the
pion decay constant. The residues in resonance pole contributions
in the vector and axial-vector correlators have the structure,
$Z_{n}^{(V,A)}=4f_{(V,A),n}^{2}m_{(V,A),n}^{2}$,
with $f_{(V,A),n}$ being defined as corresponding
decay constants.

In the scalar-pseudoscalar case it has been obtained
\cite{[12],[15]} that the residues in poles are of different order
of magnitude;

the second CSR Sum Rule  results in the estimation for splitting
between the $\sigma'$- and $\pi'$-meson masses:
$m_{\sigma'}^{2}-m_{\pi'}^{2}\simeq\frac{1}{6}m_{\sigma}^{2}$;
 and
the value $L_{8}=(0.9\pm 0.4)\cdot 10^{-3}$ from \cite{[16]}
accepts $m_{\sigma}\simeq 800$ MeV.

In the vector-axial-vector case all residues
are found to be of the same order of magnitude in contrast to the
scalar-pseudoscalar channel \cite{[19]}.
The first and the second Sum Rule is fulfilled identically
in the large-log approach. The third one takes the form:
$Z_1(m_{a_1'}^2-m_{\rho'}^2)\simeq16\pi\alpha_s<\!\bar{q}q\!>^2.
\label{3sr}$
The
structure of $Z_{\rho'}$ and $Z_{a_{1}'}$ shows that if $m_{a_{1}'}\simeq
m_{\rho'}$ then $Z_{a_{1}'}\simeq Z_{\rho'}$ and therefore
$f_{a_{1}'}\simeq f_{\rho'}$. As a consequence these residues
approximately cancel each other in Sum Rules
and after evaluating we get $f_{\rho}\approx 0.15$ and $f_{a}\approx
0.06$ to be compared with the experimental values \cite{[16]}
$f_{\rho}=0.20\pm 0.01$,
$f_{a}=0.10\pm
0.02$. We have also a
reasonable prediction for the chiral constant $L_{10}$
for the $\rho, a_{1}$-mesons and their first excitations (n=2) one gets
$L_{10}=
\approx -6.0\cdot 10^{-3}$,
which is consistent with that one \cite{[16]} from hadronic
$\tau$ decays:
$L_{10}=-(6.36\pm 0.09)\mid_{expt}\pm 0.16\mid_{theor})\cdot 10^{-3}$.
It is worth to mention also that within the four-resonance ansatz
(n=2) and using two first Weinberg sum rules one obtains the estimation
of electromagnetic pion-mass difference $\Delta m_{\pi}^{4}\mid_{em} \simeq
(3.85\pm 0.16)$ MeV, (see \cite{[17]}) which improves the agreement between theoretical
predictions and the experimental value of $\Delta m_{\pi}\mid_{em}^{expt} \simeq
(4.42\pm 0.03)$ MeV.
\section{Summary}

Let us summarize the results presented in this talk.
\begin{enumerate}
\item[(i)] The mass of the second axial-vector particle with $I=1$
is predicted. It is comparable with the mass of the vector
counter-partner: $m_{a_{1}'}=1465\div 1850 MeV$ and the most
plausible value of the mass difference is
$m_{a_{1}'}-m_{\rho'}\approx 30 MeV$;
\item[(ii)] The estimation on the mass of the $\sigma$-meson
does not contradict to existing experimental data \cite{eid}:
$m_{\sigma} \simeq 800$ MeV;
\item[(iii)] The couplings $f_{\rho},f_{a}$ and the chiral constant
$L_{10}$ as well as the electromagnetic pion-mass difference
$\Delta m_{\pi}^{4}\mid_{em}$
\cite {[17]}
are evaluated from CSR Sum Rules
as matching rules for QNJLM to QCD at
intermediate energies.
\end{enumerate}

Finally we would like to mention that the QNJL Models can be used to
describe Higgs particles in extensions of the Standard Model,
see \cite{[15],[18]}.

\section*{Acknowledgements}

We express our gratitude to the organizers of the International
Symposium "MENU 2004" in Beijing for hospitality and financial
support. This work is supported by Grant RFBR 04-02-26939, by Grant
INFN/IS-PI13 and the Program "Universities of Russia".


\end{document}